\documentclass{jltp}
\usepackage{epsfig}

\def\bi{{\bf i}}
\def\bj{{\bf j}}
\def\bk{{\bf k}}

\def\b0{{\bf 0}}
\def\cG{{\cal G}}

\def\cT{{\cal T}}

\def\bra{\langle}
\def\ket{\rangle}
\def\up{\uparrow}
\def\down{\downarrow}

\def\eps{\epsilon}

\def\om{\omega}
\def\sg{\sigma}
\def\Sg{\Sigma}

\title{Dynamical mean-field theory for the normal phase \\ 
       of the attractive Hubbard model\footnote{Dedicated to 
       P.~W\"olfle on the occasion of his 60th birthday}}

\author{M.\ Keller$^1$, W.\ Metzner$^1$, and U.\ Schollw\"ock$^2$}

\address{$^1$Theoretische Physik C, Technische Hochschule Aachen, 
  D-52056 Aachen, Germany \\
  $^2$Sektion Physik, Universit\"at M\"unchen, D-80333 M\"unchen,
  Germany}

\runninghead{M.\ Keller {\textit et al.}}{DMFT for the attractive
 Hubbard model}

\begin{document}

\maketitle


\begin{abstract}
We analyze the normal phase of the attractive Hubbard model
within dynamical mean-field theory. 
We present results for the pair-density, the spin-susceptibility,
the specific heat, the momentum distribution, and for the 
quasi-particle weight.
At weak coupling the low-temperature behavior of all quantities 
is consistent with Fermi liquid theory. 
At strong coupling all electrons are bound in pairs, which leads
to a spin gap and removes fermionic quasi-particle excitations.
The transition between the Fermi liquid phase and the pair phase
takes place at a critical coupling of the order of the band-width
and is generally discontinuous at sufficiently low temperature.

\end{abstract}

\section{INTRODUCTION}

The size of Cooper pairs in high-temperature cuprate superconductors 
is not much bigger than the average distance of conduction electrons
in these materials.\cite{Gin} 
This experimental fact has dramatically increased the interest in 
electronic model systems where attractive interactions can lead to 
bound electron pairs of arbitrary size, between the
BCS-limit of very large Cooper pairs and the opposite Bose limit,
where the pairs are smaller than the average particle 
distance.\cite{Ran95}
Already in 1980 Leggett\cite{Leg} pointed out that the 
superconducting BCS ground state at weak coupling evolves
smoothly into a Bose condensate state at strong coupling, as
a function of increasing interaction strength. 
Nozi\`eres and Schmitt-Rink\cite{NS} considered the BCS-Bose
Crossover at finite temperatures and argued that also the transition 
temperature $T_c$ between the normal and superconducting (or 
superfluid) state should evolve continuously.

There has also been much interest in possible non-Fermi liquid
behavior of the {\em normal}\/ phase of electron systems with 
attractive interactions. 
The Hubbard model for lattice electrons with a purely local
attractive interaction\cite{MRR} has become a prototype model in 
this context.
A T-matrix calculation by Fr\'esard et al.\cite{FGW} for the
attractive Hubbard model showed convincingly that Fermi liquid 
theory governs the normal phase for relatively weak coupling 
strength even in two dimensions, except for very low density.
Only very close to $T_c$ deviations from Fermi liquid behavior
due to superconducting fluctuations occur at weak 
coupling.\cite{RM}

Quantum Monte Carlo (QMC) simulations of the 
two-dimensional\cite{RTMS,SPSBM} and three-dimensional\cite{San} 
attractive Hubbard model have established the formation of a spin 
gap and a gap in the single-particle excitation spectrum in the
normal phase at sufficiently strong coupling. 
Approximate theories beyond the T-matrix \cite{AT} have 
produced quite strong pseudogap behavior at intermediate
interaction strength in two dimensions.\cite{KAT}
These results have been related to pseudogap phenomena in 
underdoped cuprate superconductors.\cite{Ran98}

In this article we analyze the pair formation and related
phenomena in the normal phase of the attractive Hubbard model
within the dynamical mean-field theory (DMFT).\cite{GKKR}
This approximation becomes exact in the limit of infinite
lattice dimension.\cite{MV}
We solve the mean-field equations numerically at finite 
temperature.
The results show that the normal state is a Fermi liquid at
weak coupling and a non-Fermi liquid state characterized by
bound electron pairs, the absence of fermionic quasi-particles
and a spin gap at strong coupling, in qualitative agreement
with the QMC studies of finite two- and three-dimensional
systems.\cite{RTMS,SPSBM,San}
At very low temperatures the transition between the Fermi 
liquid and the normal paired state is discontinuous, if the
superconducting instability is suppressed.
A short account of this work has already appeared.\cite{KMS01}
Here we give more details on the method and present more low 
temperature data as well as new results for physical quantities
not discussed previously.
Our analyis is very nicely complemented by a very recent 
computation of spectral properties of the DMFT solution at zero
temperature by Capone et al.\cite{CCG}

In Sec.\ 2 we introduce the attractive Hubbard model and 
discuss some of its elementary properties. 
In Sec.\ 3 we motivate and describe the DMFT, with some
details on its evaluation.
Sec.\ 4 is dedicated to the presentation and interpretation 
of results. 
Most results have been obtained for quarter-filling, but we also 
present some results for half-filling and filling factor one 
eighth, to show how the pairing transition depends on density.
In Sec.\ 5 we summarize the results and discuss deficiencies
of the DMFT.

\section{ATTRACTIVE HUBBARD MODEL}

The Hubbard model for lattice electrons with a nearest neighbor
hopping amplitude $-t$ and a local interaction $U$ ist given by
\begin{equation}\label{Hubbard}
 H = -t \sum_{\bra \bi,\,\bj \ket} \sum_{\sg} 
 t_{\bi\bj} \, c^{\dag}_{\bi\sg} \, c_{\bj\sg} 
 \, + \, U \, \sum_{\bj} n_{\bj\up} \, n_{\bj\down} \; ,  
\end{equation}
where $c^{\dag}_{\bi\sg}$ and $c_{\bi\sg}$ are the usual creation 
and annihilation operators for fermions with spin projection
$\sg \in \{\up,\down\}$ on a lattice site $\bi$, and
$n_{\bj\sg} = c^{\dag}_{\bj\sg} c_{\bj\sg}$.
The first lattice sum is restricted to nearest neighbors $\bi$
and $\bj$.
For the {\em attractive}\/ Hubbard model\cite{MRR} the coupling 
constant $U$ is negative. 

The attractive Hubbard model is expected to be a superconductor
below a certain critical temperature $T_c(U,n) > 0$ for all 
$U < 0$ at any average density $n$, if the lattice dimensionality 
is above two.\cite{MRR}
At half-filling ($n=1$) the usual $U(1)$ gauge symmetry becomes 
a subgroup of a larger $SO(3)$ symmetry, and the superconducting
order parameter mixes with charge density order.
In two dimensions one expects a Kosterlitz-Thouless phase at low 
temperatures for all $U < 0$ and $n \neq 1$, with a finite 
superfluid density and quasi long-range order.
At half-filling the non-Abelian $SO(3)$ symmetry excludes the 
possibility of a Kosterlitz-Thouless phase.

In the weak coupling limit $U \to 0$ and dimensions $d > 2$ the 
attractive Hubbard model can be reasonably treated by BCS mean-field 
theory.\cite{NS,MRR}
In the strong coupling limit $U \to -\infty$ the low energy 
sector of the model (excitation energies $\ll |U|$) can be 
mapped onto an effective model of hard core lattice bosons with
a hopping amplitude of order $t^2/U$ and a repulsive nearest 
neighbor interaction of the same order.\cite{NS,MRR} 
These bosons undergo Bose condensation in $d>2$ dimensions and
a Kosterlitz-Thouless transition in two dimensions (for $n \neq 1$) 
at a critical temperature of order $t^2/|U|$.

For nearest neighbor hopping on a bipartite lattice the particle-hole 
transformation of spin-$\up$ fermions
\begin{equation}\label{phtrafo}
 c_{\bj\up} \mapsto \eta_{\bj} \, c^{\dag}_{\bj\up} \; , \quad 
 c^{\dag}_{\bj\up} \mapsto \eta_{\bj} \, c_{\bj\up} \, ,
\end{equation}
where $\eta_{\bj} = 1 \; (-1)$ for $\bj$ on the A-sublattice
(B-sublattice), maps the attractive Hubbard model at density
$n$ onto a repulsive Hubbard model at half-filling with a
finite average magnetization $m = 1-n$.\cite{MRR}
This relation is useful to compare with results known for the 
repulsive Hubbard model.

\section{DYNAMICAL MEAN-FIELD THEORY}

We have solved the attractive Hubbard model within DMFT.\cite{GKKR}
In contrast to other (simpler) mean-field approaches, DMFT provides 
an exact solution of the model in the limit of infinite lattice 
dimensionality,\cite{MV} since it captures local fluctuations
exactly. 

At weak coupling DMFT incorporates the complete BCS physics, since 
it contains the Feynman diagrams contributing to the BCS mean-field 
theory. 
At strong coupling, where the attractive Hubbard model maps to the
hard core Bose gas, DMFT reduces to the standard mean-field theory 
of the hard core Bose gas.\cite{fn1}
Hence, Bose-Einstein condensation of preformed pairs is obtained at 
a critical temperature of order $t^2/|U|$ at large $|U|$,
\begin{eqnarray}\label{Tcbose}
 T_c = z \, \frac{2t^2}{|U|} \, \frac{n-1}{\ln\frac{n}{2-n}}
\end{eqnarray}
where $z$ is the coordination number of the lattice.

Within DMFT the fluctuating environment of any lattice site is
replaced by a local but dynamical effective field 
$\cG_0(\tau,\tau')$.\cite{GKKR}
The mean-field equations involve the calculation of the propagator
$G(\tau,\tau') = 
 - \bra \cT c_{\sg}(\tau) \, c^{\dag}_{\sg}(\tau') \ket$ 
of an effective single-site Hubbard model coupled to the dynamical 
field $\cG_0$, 
and a self-consistency condition relating $G$ to the 
local propagator of the full lattice system.
The effective single-site action reads
\begin{eqnarray}\label{S}
 S = \sum_{\sg} \int_0^{\beta} \!d\tau \int_0^{\beta} \!d\tau' \,
 \cG_0^{-1}(\tau,\tau') \, c^{\dag}_{\sg}(\tau) \, c_{\sg}(\tau')
 - U \int_0^{\beta} \!d\tau \, n_{\up}(\tau) \, n_{\down}(\tau)
\end{eqnarray}
where $\beta = 1/T$ and $\cG_0^{-1}$ is the inverse of $\cG_0$
(in the sense of a linear integral operator).

The lattice structure enters only via the bare density of
states (DOS) into the self-consistency condition, as long as the 
translation invariance of the lattice is not broken. 
We have used the particularly simple self-consistency 
equations\cite{GKKR}
\begin{equation}
 \cG_0^{-1}(i\om_n) = i\om_n + \mu 
  - (\eps_0/2)^2 \, G(i\om_n)
\end{equation}
corresponding to a half-ellipse shaped density of states 
$D_0(\eps) = \frac{2}{\pi\eps_0^2} \sqrt{\eps_0^2-\eps^2}$.
Here $G(i\om_n)$ and $\cG_0(i\om_n)$ are the Fourier transforms
of $G(\tau,\tau')$ and $\cG_0(\tau,\tau')$, respectively.
Any other bounded DOS would yield qualitatively similar results.
A simple lattice system yielding the half-ellipse 
$D_0(\eps)$ is the Bethe lattice with a nearest neighbor hopping 
amplitude $t=t^*/\sqrt{z}$ in the infinite coordination number
limit $z \to \infty$, where $\eps_0/2 = t^*$.
In the following we will set $\eps_0/2 = 1$ such that the bare
bandwidth is $W_0 = 4$.

Susceptibilities such as the pairing and the spin susceptibility 
can also be computed from expectation values of operator products 
within the effective single-site problem. 
The DMFT equations can also be extended to superconducting
or other symmetry broken phases.\cite{GKKR}
In this work we focus however on normal state properties.  

The effective single-site problem appearing in the DMFT for the 
Hubbard model can be related to the Anderson model of a single
Hubbard impurity coupled via a hybridization term to a bath of
non-interacting conduction electrons.\cite{GKKR} Integrating
out the conduction electrons of this model yields an effective
action of the form (\ref{S}). The Weiss field is determined by 
the parameters of the Anderson model as
\begin{eqnarray}
 \cG_0^{-1}(i\om_n) = i \om_n + \mu -
 \int_{-\infty}^{\infty} d\om \, \frac{\Delta(\om)}{i\om_n - \om}
\end{eqnarray}
where the hybridization spectral density $\Delta(\om)$ is
given by the conduction band energy levels $\eps_l$ and the
corresponding hybridization matrix elements $V_l$ of the
Anderson model as
\begin{eqnarray}
 \Delta(\om) = \sum_{l,\sg} V_l^2 \, \delta(\om - \eps_l)
\end{eqnarray}

The effective single-site problem (\ref{S}) cannot be solved 
analytically.
We have solved it numerically by discretizing the imaginary time 
interval $[0,\beta]$ into $L$ time slices of size 
$\Delta\tau = \beta/L$ and computing expectation values via the 
negative $U$ analogue of the Hirsch-Fye algorithm.\cite{HF}
The evaluation of the discretized path integral is reduced 
via a discrete Hubbard-Stratonovich transformation to a sum over 
Ising-spin configurations with $L$ spins in this algorithm.
The $2^L$ different configurations have been summed exactly for 
$L \leq 24$ (intermediate and high temperatures) and by a 
Monte-Carlo routine with importance sampling for $L > 24$ 
(low temperatures).
Most results have been computed for $\Delta\tau = 0.2$.
A $\Delta\tau$-extrapolation to $\Delta\tau \to 0$ has been 
performed in cases where significant $\Delta\tau$-dependences
were observed.

\section{RESULTS}

We now present and discuss results for the normal phase of the
attractive Hubbard model as obtained from our DMFT calculation.
Most of the results have been obtained at quarter-filling 
($n=1/2$). 
We do not expect that the results depend qualitatively on the
density in the attractive Hubbard model, as long as $n$ is finite.
Only the particle-hole symmetric half-filled case ($n=1$) is special 
due to its larger symmetry group.
Quarter-filling is well below half-filling but still high enough 
to see collective many-body effects, which are not obtained in the 
low-density limit.
Note that the chemical potential is known analytically only at
half-filling (where $\mu = U/2$), while it has to be determined
numerically in a self-consistency loop for $n \neq 1$. 
For $n<1$ the chemical potential is a monotonously decreasing 
function of temperature.

Results for the critical temperature $T_c(U)$ for the onset of 
superconductivity at quarter-filling have been presented in our
recent letter\cite{KMS01} and will not be reproduced here.
At half-filling $T_c(U)$ has been computed already much earlier
by Freericks et al.\cite{FJS} By virtue of the particle-hole
symmetry at half-filling the critical temperature is equal to the 
N\'eel temperature of the repulsive Hubbard model in that case. 
$T_c(U)$ is exponentially small at weak coupling and approaches
the Bose-limit (\ref{Tcbose}) for strong coupling, as expected.
At all coupling strengths our numerical results for $T_c(U)$ 
vary smoothly as a function of $U$, as expected from the 
arguments of Nozi\`eres and Schmitt-Rink.\cite{NS}

In the following we concentrate on physical properties of the
{\em normal}\/ phase. 
We ignore the superconducting instability and study normal state
solutions of the DMFT equations also below $T_c$. 
Of course these solutions do not minimize the free energy, but
they could be stabilized by the field energy of a sufficiently
strong external magnetic field.

\subsection{Fermi liquid and pair phase}

At weak coupling the normal state of the system is a Fermi liquid
with fermionic quasi-particle excitations. 
Besides numerical evidence (see below) this follows\cite{fn2} 
from the analyticity of weak coupling perturbation theory for the 
effective single-site problem. 
By contrast, at sufficiently strong coupling $|U| \gg W_0$ and 
zero temperature all particles should be bound in pairs, because a
small kinetic energy cannot overcome a finite binding energy. 
At low finite temperatures $T \ll |U|$ only an exponentially 
small fraction of pairs dissociates.
Our DMFT results at strong coupling are indeed characterized
by the absence of fermionic low-energy excitations and a spin
gap associated with the binding in singlet pairs.

A direct measure for local pair formation is the 
{\em local pair density}, that is the density of doubly occupied 
sites $n_d = \bra n_{\bj\up} \, n_{\bj\down} \ket \:$.
For an uncorrelated state the density of doubly occupied sites
is simply the product of the average density of up and down spin 
particles, i.e.\ $n_d^0 = n_{\up} \, n_{\down} = (n/2)^2$.
An attractive interaction enhances $n_d$.  
In the limit of infinite attraction all particles are bound as 
local pairs such that $n_d \to n/2$. 
In Fig.\ 1 we show results for $n_d(T)$ for various $U$.
\begin{figure}[htb]
\begin{center}
\vspace{1cm}
\epsfig{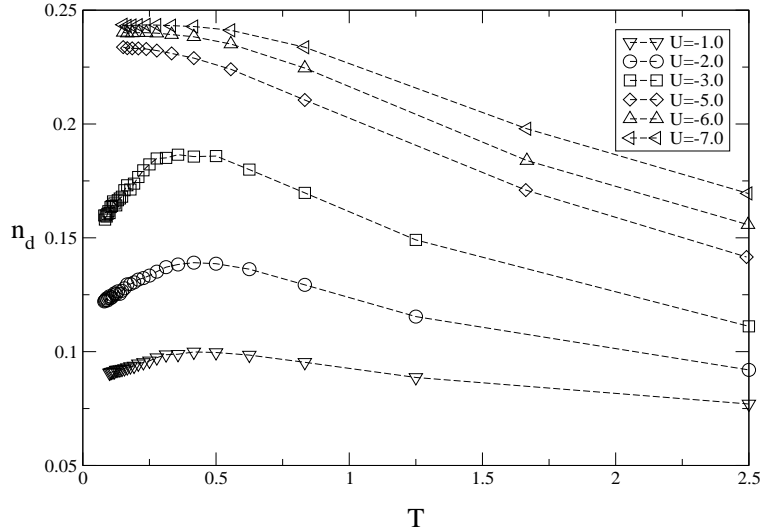} 
\vspace{5mm}
\caption{Density of doubly occupied sites $n_d$ as a function of
 temperature for various coupling strenghts $U$ at quarter-filling.} 
\end{center}
\end{figure} 
For $T \to \infty$ the density of doubly occupied sites tends to
$(n/2)^2$, corresponding to an uncorrelated state.
For decreasing temperature $n_d(T)$ first increases as a 
consequence of the attractive interaction. 
For small or moderate $U$, however, $n_d(T)$ slightly decreases 
again at low temperatures. This effect, which has also been 
obtained in a combined DMFT + TMA calculation,\cite{KMS99} can 
be attributed to the kinetic energy, which tends to dissociate 
pairs if the attraction is not too strong.
Note that in the pairing regime for stronger $U$ the upturn in
$n_d(T)$ at low temperatures is missing. The kinetic energy is
not able to unbind pairs any more.
For the largest $|U|$ values $n_d(T)$ becomes very flat at low 
temperatures, which indicates the presence of an energy gap for 
excitations. 
A completely analogous (particle-hole transformed) behavior has 
been found in the DMFT solution of the repulsive Hubbard model at
half-filling.\cite{GKKR}

The binding of all electrons in singlet pairs in the pairing state 
at strong coupling leads to a {\em spin gap}\/, which can be 
observed most directly in the {\em spin susceptibility}.
In Fig.\ 2 we show our DMFT results for the temperature dependence 
\begin{figure}[htb]
\begin{center}
\vspace{1cm}
\epsfig{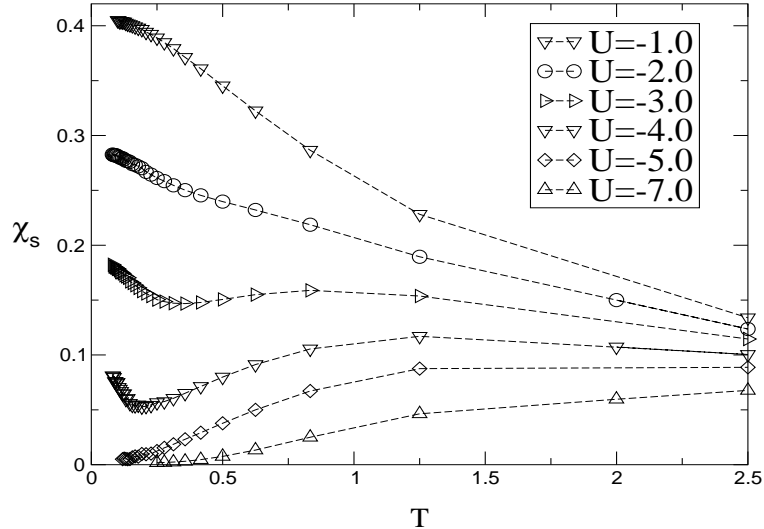} 
\vspace{5mm}
\caption{Spin susceptibility $\chi_s$ as a function of temperature
 for various coupling strengths $U$ at quarter-filling.} 
\end{center}
\end{figure} 
of the spin susceptibility $\chi_s$, for various coupling strengths.
The results have been obtained by computing the spin-spin correlation
function, which yields much more accurate data than the alternative
route via numerical diffentiation of the magnetization in a
small external magnetic field. 
Due to a rather strong $\Delta\tau$-dependence of the data
a $\Delta\tau$-extrapolation had to be performed here.
For a weak attraction the spin susceptibility increases
monotonously for lower temperatures and then saturates at a 
finite value for $T \to 0$, as expected for a Fermi liquid.
For strong coupling, however, $\chi_s$ decreases rapidly at low 
temperatures, as expected for a system where spin excitations are 
gapped. 
This gap, which has also been seen in QMC simulations of the
two-dimensional\cite{RTMS,SPSBM} and three-dimensional\cite{San}
Hubbard model, is clearly due to the binding energy of singlet pairs 
in the non-Fermi liquid state forming at strong coupling.
For a moderate attraction, {\em pseudogap}\/ behavior seems to set 
in at intermediate temperatures, but for small $T$ the susceptibility 
increases again and finally tends to a non-zero value.
In our earlier work\cite{KMS01} we attributed this behavior to 
the presence of a narrow quasi-particle band in the system, 
similar to the one known for the repulsive model near the Mott 
transition.\cite{GKKR} 
The very recent results for the spectral function at 
quarter-filling by Capone et al.\cite{CCG} indeed confirm this 
expectation.

Results for the {\em specific heat}\/ $c_v(T)$ are shown in 
Fig.\ 3. 
\begin{figure}[htb]
\begin{center}
\vspace{1cm}
\epsfig{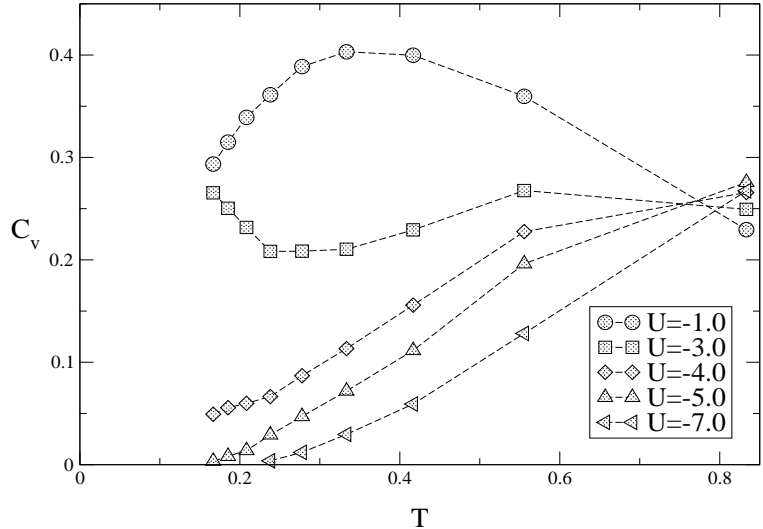} 
\vspace{5mm}
\caption{Specific heat $c_v$ as a function of temperature
 for various coupling strengths $U$ at quarter-filling.} 
\end{center}
\end{figure} 
The specific heat is obtained by numerically differentiating the
results for the internal energy $E(T)$, which unfortunately
amplifies the statistical fluctuations in the Monte-Carlo data
quite drastically. 
We therefore show only results at relatively high temperature,
where the Ising spin configurations in the Hirsch-Fye algorithm
can still be enumerated exactly.
Nevertheless one can clearly see that the specific heat 
exhibits activated behavior with an energy gap for large $|U|$.
In the Fermi liquid regime for small $|U|$ the specific heat 
must vanish linearly for $T \to 0$, but this asymptotic behavior
sets in only at rather low temperatures, even for very moderate
$|U|$. 
In the strongly correlated Fermi liquid regime for intermediate
$U$ the additional energy scale observed already in the
spin susceptibility is also visible in the specific heat:
$c_v(T)$ first decreases at intermediate temperatures $T > 0.2$, 
before increasing again (at $U=-3$) or at least saturating 
(at $U=-4$) at a lower scale. 
Ultimately $c_v(T)$ must of course vanish linearly in the zero 
temperature limit for $U$-values in the Fermi liquid regime. 

The existence of an energy gap in the specific heat implies 
that there are no low energy excitations at all in the pair
phase. 
We emphasize that this is an artefact of the DMFT, which does
not take into account the contributions from the low-energy 
bosonic degrees of freedom to the specific heat. 
To see how this comes about let us consider the limit of 
high lattice dimensionality (coordination number $z \to \infty$), 
with the scaling of the hopping amplitude 
$t = t^*/\sqrt{z} \;$.\cite{MV} 
The DMFT solves the Hubbard model exactly in that limit.
The scaling of $t$ has been chosen such that the average
kinetic energy of the electrons has a finite limit for 
$z \to \infty$, but the effective hopping amplitude $t_b$ of the 
composite bosons forming for large $|U|$ is proportional to 
$t^2/|U| = z^{-1}\, (t^*)^2/|U|$, such that the average 
kinetic energy of the bosons is suppressed by a factor $1/z$ in 
the limit $z \to \infty$. 
Only in a Bose condensate the kinetic energy contributions
of all bonds on the lattice add up coherently to a finite total 
kinetic energy.
The situation is completely analogous to that for the repulsive 
Hubbard model, where the spin exchange energy $J$ is suppressed 
as $1/z$, and thus yields a finite energy gain only via magnetic
ordering in the large $z$ limit.\cite{GKKR}

%
%

The momentum distribution function 
$n_{\bk} = \bra c^{\dag}_{\bk\sg} c_{\bk\sg} \ket$ 
also behaves differently in the Fermi liquid phase and the pair 
phase, respectively.
Within DMFT, where the self-energy is momentum-independent,
$n_{\bk}$ depends only via the single particle energy $\eps_{\bk}$
on $\bk$, defining thus an {\em ''energy distribution function''}\/
$n(\eps)$.
An analogous definition works also for the Bethe lattice or 
other systems where single particle states are labeled by other 
quantum numbers than momentum.
In Fig.\ 4 we show results for $n(\eps)$ for different choices
of $U$. 
\begin{figure}[htb]
\begin{center}
\vspace{1cm}
\epsfig{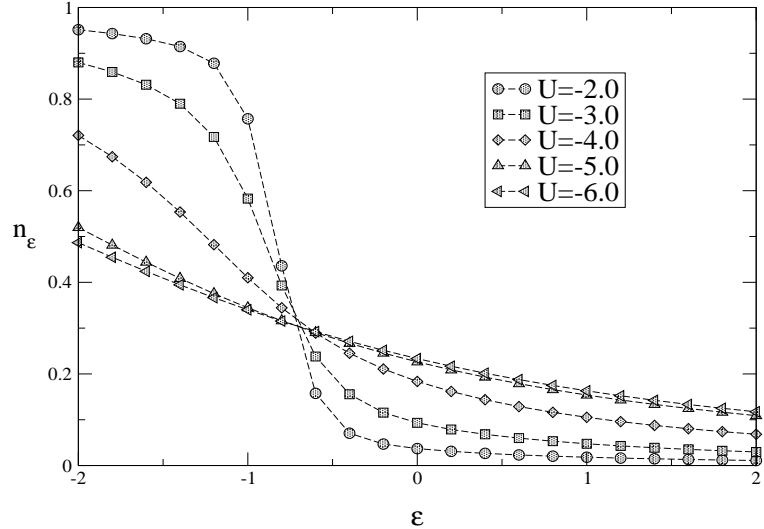} 
\vspace{5mm}
\caption{Energy distribution function $n(\eps)$ for various coupling
 strengths $U$ at quarter-filling; the temperature is fixed at 
 $T=0.08$.}
\end{center} 
\end{figure}
One can clearly see that $n(\eps)$ changes curvature
near the Fermi level, if $U$ is in the Fermi liquid regime,
while the energy distribution function becomes convex in the
pair phase. Hence there is nothing like a Fermi surface in the
pair phase. 

\subsection{Phase transition}

We now analyze the phase transition between the Fermi liquid
at weak coupling and the pair phase at strong coupling in 
more detail. 
Since the Fermi liquid state is qualitatively different from 
the pair state, there has to be a sharply defined 
{\em pairing transition}\/ at some critical attraction $U_c$ at 
least in the ground state.
At finite temperature one may expect either a genuine phase
transition or, alternatively, a smooth (possibly very steep) 
crossover.
At half-filling the attractive Hubbard model is equivalent
to the spin-symmetric repulsive model, for which the existence
of a {\em first order}\/ phase transition between the Fermi 
liquid and the paramagnetic Mott phase at sufficiently low finite 
temperatures is well established.\cite{GKKR}
The first order transition line in the $(U,T)$-plane is 
embedded in a region where two different solutions of the DMFT 
equations, with Fermi liquid and Mott insulator properties, 
respectively, coexist.
The particle-hole binding characterizing the Mott phase 
translates into particle-particle binding in the attractive
case.
Away from half-filling the attractive Hubbard model maps to
the repulsive model at half-filling with a finite magnetization.
For that model Laloux et al.\cite{LGK} have solved the DMFT
equations with an exact diagonalization algorithm, finding
coexisting solutions also away from the spin symmetric case. 
The results of their work imply that at sufficiently low 
temperatures a first order pairing transition occurs in the 
attractive Hubbard model also away from half-filling.

To see how the Fermi liquid phase breaks down upon increasing 
the attraction strength, we have computed the renormalization 
factor 
\begin{eqnarray}
 Z(T) = \left[ 1 - \frac{\Sg(i\omega_0)}{i\omega_0}
 \right]^{-1} 
\end{eqnarray}
where $\Sg$ is the self-energy and $\om_0 = \pi T$ the smallest 
(positive) Matsubara frequency at temperature $T$.
In Fig.\ 5 we plot $Z(T)$ as a function of $T$ for various $U$ 
at quarter-filling.
\begin{figure}[t]
\begin{center}
\vspace{1cm}
\epsfig{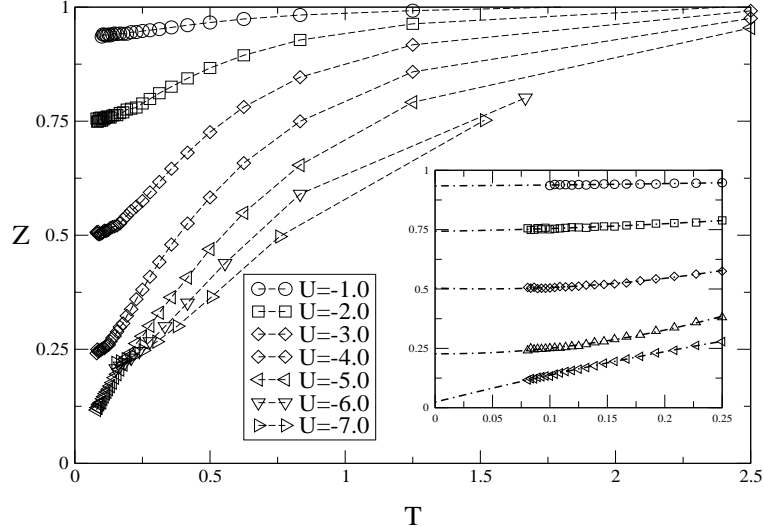} 
\vspace{2mm}
\caption{Renormalization factor $Z$ as a function of temperature
 for various $U$ at quarter-filling; 
 the inset shows low-temperature data and their quadratic fit.} 
\end{center}
\end{figure}
The inset shows the low-temperature behavior of $Z(T)$ and 
its quadratic extrapolation to $T \to 0$.
The resulting extrapolated values of $Z$ at $T=0$ are presented
in Fig.\ 6, where the corresponding results at half-filling are
also shown for comparison (see inset).
\begin{figure}[htb]
\begin{center}
\vspace{1cm}
\epsfig{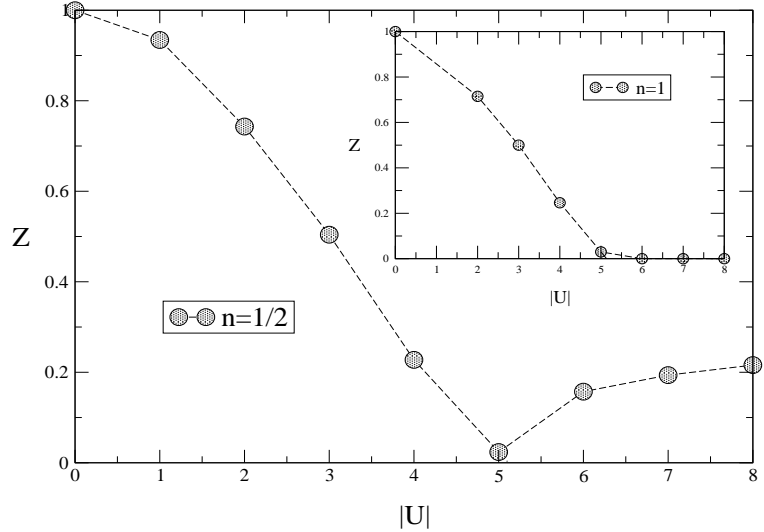} 
\vspace{2mm}
\caption{Results of the zero-temperature extrapolation of the 
 $Z$-factor as a function of $U$ at quarter-filling; the inset shows
 the extrapolated $Z$-factor at half-filling.} 
\end{center}
\end{figure}
At half-filling we find that $Z$ decreases continuously as a 
function of increasing $|U|$ in the Fermi liquid phase until
it vanishes, and remains zero in the pair phase, as is well
established for the equivalent repulsive Hubbard model.\cite{GKKR}
At quarter-filling the behavior is very different: 
$Z$ first decreases as a function of $|U|$ in the Fermi liquid
regime, goes through a minimum, and then increases again in
the pair phase. 
Obviously $Z$ is finite also in the pair phase for electron
densities away from half-filling.

In the Fermi liquid phase $Z$ has a multiple physical meaning:
$Z$ is the spectral weight for quasi-particles, the Fermi edge 
discontinuity in the momentum distribution function and, within 
DMFT, also the inverse mass renormalization.
This meaning is of course lost in the bound pair state.
However, the finiteness of $Z$ does not imply that the system is 
a Fermi liquid. Obviously it does not even imply that there are
low-energy excitations in the system.
A simple calculation shows that $Z$ is finite even in the atomic 
limit $t=0$ for $n \neq 1$, where the system is obviously not a 
Fermi liquid.

Our numerical data suggest that the minimum value of $Z$ (as a
function of $U$) is small but finite at quarter-filling, 
but there remains an uncertainty due to the extrapolation 
from finite to zero temperature. 
Most recently Capone et al.\cite{CCG} have clarified this point 
by solving the DMFT equations via an exact diagonalization 
algorithm\cite{CK} which works directly at zero temperature.
They found that the minimal $Z$ at quarter-filling is indeed 
tiny but finite.

It is instructive to consider lower densities for comparison.
In the extreme low density limit $n \to 0$ the bound pair state 
is stable once the attraction exceeds the threshold for two-particle 
binding $U_c^0$. For $|U| < |U_c^0|$ no bound states exist,
and the particles move essentially freely, due to the low
density, and $Z$ is almost one even close to the pairing 
transition.
In Fig.\ 7 we show results for $Z$,
as obtained from a zero temperature extrapolation of our
finite temperature data at filling factor one eighth ($n=1/4$).
\begin{figure}[htb]
\begin{center}
\vspace{1cm}
\epsfig{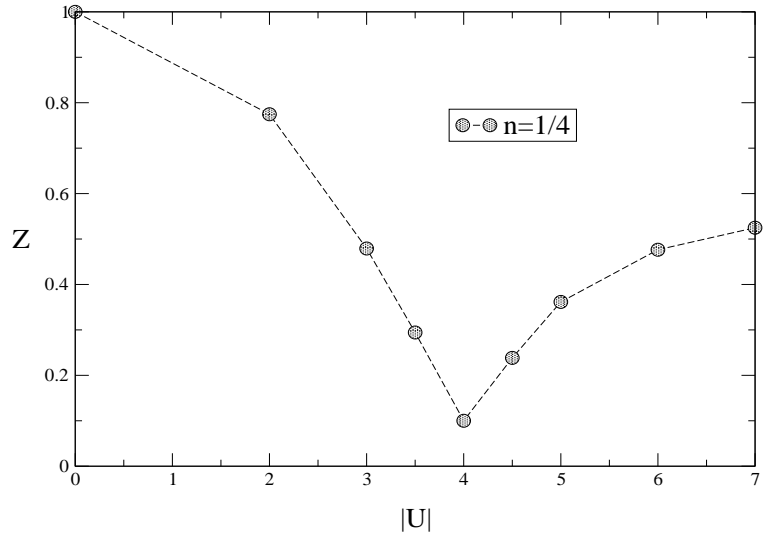} 
\vspace{2mm}
\caption{Results of the zero-temperature extrapolation of the 
 $Z$-factor as a function of $U$ at filling one eighth ($n=1/4$).} 
\end{center}
\end{figure}
One can see two trends, compared to quarter-filling:
the minimum shifts towards smaller $|U|$, moving thus closer
to the still smaller $|U_c^0|$, and the minimum value is now
significantly higher, such that its finiteness can be concluded
with more confidence from our data.

That the behavior of the $Z$-factor in the symmetric case is
different from the generic scenario is plausible also from the 
following {\em ''Kondo''}\/ point of view, which is most easily 
visualized for the repulsive model.
In the spin symmetric case (no magnetization) the spin degree
of freedom of the impurity atom in the effective single impurity
Anderson model is degenerate. 
In the strongly correlated Fermi liquid the effective Anderson
model is in the Kondo regime, and the narrow quasi-particle peak 
in the (interacting) density of states is associated with the
Kondo resonance of the Anderson model.\cite{GKKR}
A magnetic field lifts the spin degeneracy and thus destroys
the Kondo resonance at least at low energy scales.
Hence it is hard to believe that the symmetric transition scenario, 
where the quasi-particle peak vanishes continuously by 
becoming increasingly narrow, survives in the asymmetric case.
In the attractive case it is the degeneracy between empty
and doubly occupied sites which plays the role of the spin
degeneracy for the repulsive model.

The above-mentioned existence of coexisting Fermi liquid and pair 
solutions, which has been obtained in particle-hole transformed
form already by Laloux et al.\cite{LGK} for the repulsive Hubbard
model, has been confirmed in detail by Capone et al.\cite{CCG} 
Using the Hirsch-Fye algorithm it is not easy to access 
sufficiently low temperatures to reach the coexistence region
in the $(U,T)$ plane, especially away from half-filling, where
the computation of the chemical potential requires additional
self-consistency loops.
Nevertheless we have found some cases where a Fermi liquid and
a pair solution coexists. An example at quarter-filling is
shown in Fig.\ 8. 
\begin{figure}[htb]
\begin{center}
\vspace{1cm}
\epsfig{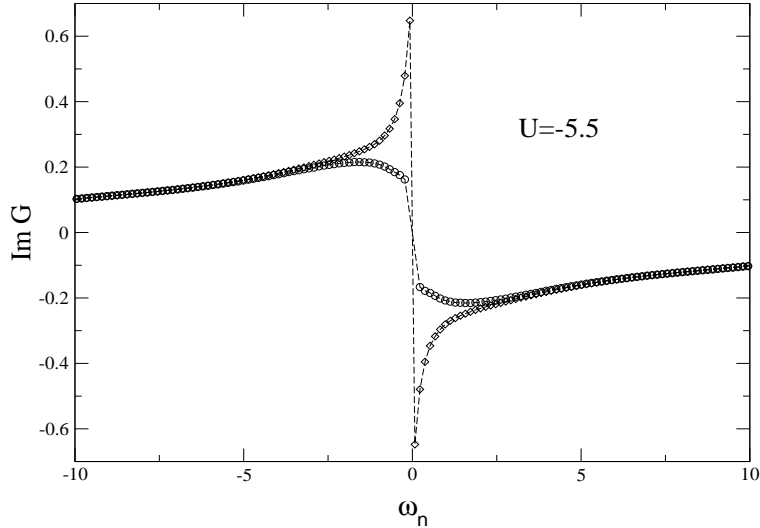} 
\vspace{2mm}
\caption{Imaginary parts of the local propagator $G$ from two
 coexisting solutions of the DMFT equations at $U = -5.5$ and
 $T = 0.023$. 
 The more singular solution belongs to the Fermi liquid phase, 
 the other to the pair phase.} 
\end{center}
\end{figure} 
Both Green functions are fully converged
self-consistent solutions of the DMFT equations, with a time
discretization $\Delta\tau = 0.35$ and $140000$ Monte Carlo 
sweeps in each iteration.

The actual phase transition takes place inside the coexistence 
region. 
To determine the transition line in the $(U,T)$-plane one would 
have to compare the free energies of the two solutions.
Since the $Z$-factor decreases as a function of $|U|$ in the 
Fermi liquid solution and increases in the pair solution, it 
is clear that $Z$ is minimal at the transition point.

\section{CONCLUSIONS}

In summary, within DMFT two distinct normal low temperature 
phases are found for the attractive Hubbard model at arbitrary
filling factor: a Fermi liquid phase at weak coupling and a
singlet pair phase characterized by a spin gap and the absence 
of fermionic quasi-particles at strong coupling.
The transition between the two phases is generally first order
and occurs at an intermediate critical coupling $U_c$ of the
order of the bandwidth, which is maximal at half-filling and
(most probably) minimal in the low-density limit $n \to 0$. 
Our numerical results, especially at filling factor one eighth, 
the numerical results by Capone et al.,\cite{CCG} and analytical
arguments all indicate that the Fermi liquid ground state 
disappears with a finite $Z$ at $U_c$ at electron densities 
$n \neq 1$, such that in contrast to the special half-filled 
case the quasi-particle weight disappears discontinuously at 
the pairing transition.

The DMFT provides an exact solution of the Hubbard model in
the limit of infinite lattice dimension (or coordination number).
As an approximation for the two- or three-dimensional model
it captures at least the most gross features of the normal phase,
that is Fermi liquid behavior at weak coupling and a singlet pair 
liquid with a spin gap at strong coupling. 
The instability towards a superfluid state at an exponentially
small temperature scale at weak coupling and at a scale of order 
$t^2/|U|$ is also described by DMFT.

The bosonic degrees of freedom in the pair phase are however
poorly treated by DMFT. Their kinetic energy is taken into 
account only in a Bose condensate, while in the normal phase
the pairs do not move. 
Gapless bosonic excitations, which are present in any finite 
dimension, are absent in DMFT.

Within DMFT, the development of the gap in the normal phase
is accompanied by a complete destruction of the Fermi edge in 
the momentum distribution function. 
Furthermore, there is not even a trace of pseudo gap behavior
at weak coupling. 
This is different in two dimensions, where superconducting
fluctuations lead to a small pseudo gap in the normal phase 
close to $T_c$ already at weak coupling,\cite{RM} and a rather
strong pseudo gap develops at moderate coupling strength in a
regime where the momentum distribution function still exhibits
a pronounced Fermi edge.\cite{Ran98}

\bigskip


\section*{ACKNOWLEDGMENTS}
W.M. would like to thank A.\ Georges for his kind hospitality
and valuable discussions at Ecole Normale Superieure.
Discussions with M. Capone and C. Castellani are also gratefully 
acknowledged. 
This work has been supported by the Deutsche Forschungsgemeinschaft 
under Contract No.\ Me 1255/5.



\begin{thebibliography}{99}
\bibitem{fn1} The hard core Bose gas can be mapped to an 
 antiferromagnetic Heisenberg model in an external magnetic field,
 and its mean-field theory maps to the standard mean-field theory
 of the Heisenberg model; 
 see, for example Ref.\ \onlinecite{MRR}, Sec.\ IIB.
\end{thebibliography}
\end{document}